\documentclass[runningheads,a4paper]{llncs}

\usepackage{amssymb}
\usepackage{graphicx}
\usepackage{epstopdf}
\usepackage{multirow}
\usepackage{amsmath}
\usepackage{caption}
\usepackage{array}
\usepackage{url}
\newcolumntype{C}[1]{>{\centering\let\newline\\\arraybackslash\hspace{0pt}}m{#1}}

\urldef{\mailsa}\path|{alfred.hofmann, ursula.barth, ingrid.haas, frank.holzwarth,|
\urldef{\mailsb}\path|anna.kramer, leonie.kunz, christine.reiss, nicole.sator,|
\urldef{\mailsc}\path|erika.siebert-cole, peter.strasser, lncs}@springer.com|
\newcommand{\keywords}[1]{\par\addvspace\baselineskip
\noindent\keywordname\enspace\ignorespaces#1}

\begin{document}

\mainmatter  

\title{Towards Formal Fault Tree Analysis using Theorem Proving \thanks{ The final publication is available at http://link.springer.com}}


%
%
\author{Waqar Ahmed %
\and Osman Hasan }
%

\institute{School of Electrical Engineering and Computer Science (SEECS)\\
National University of Sciences and Technology (NUST)\\
Islamabad, Pakistan \\
\email{ \{waqar.ahmad,osman.hasan\}@seecs.nust.edu.pk  }
}

%
%

\maketitle

\begin{abstract}
Fault Tree Analysis (FTA) is a dependability analysis technique that has been widely used to predict reliability, availability and safety of many complex engineering systems. Traditionally, these FTA-based analyses are done using paper-and-pencil proof methods or computer simulations, which cannot ascertain absolute correctness due to their inherent limitations. As a complementary approach, we propose to use the higher-order-logic theorem prover HOL4 to conduct the FTA-based analysis of safety-critical systems where accuracy of failure analysis is a dire need. In particular, the paper presents a higher-order-logic formalization of generic Fault Tree gates, i.e., AND, OR, NAND, NOR, XOR and NOT and the formal verification of their failure probability expressions. Moreover, we have formally verified the generic probabilistic inclusion-exclusion principle, which is one of the foremost requirements for conducting the FTA-based failure analysis of any given system. For illustration purposes, we conduct the FTA-based failure analysis of a solar array that is used as the main source of power for the Dong Fang Hong-3 (DFH-3) satellite.
\end{abstract}

\section{Introduction}
With the increasing usage of engineering systems in safety-critical domains, their dependability and failure analysis \cite{international2006iec} has become a dire need to predict their reliability, availability and safety. One of the most widely used dependability and failure analysis techniques is the Fault Tree Analysis (FTA) method \cite{roberts1987fault}. It is a graphical technique consisting of internal nodes, which are represented by gates like OR, AND and XOR, and the external nodes, that model the events which are associated with the occurrence of faults in sub-systems or components of the given system. The generic nature of these gates and events allows us to construct an efficient and accurate fault tree (FT) model for any given system. This FT can in turn be used to investigate the potential causes of a fault occurrence in a system and the calculation of minimal number of events that contribute towards the occurrence of a $top$ $event$, i.e., a critical event, which can cause the whole system failure upon its occurrence. Some noteworthy applications of FTA include the failure analysis of transportation systems \cite{huang2000fuzzy}, healthcare systems \cite{hyman2008fault} and aerospace systems \cite{wu2011reliability}.

Traditionally, FTA is carried out by using paper-and-pencil proof methods, computer simulations and computer algebra systems. The first step in the paper-and-pencil proof methods is the construction of the FT of the given system on a paper.
This is followed by the identification of the Minimal Cut Set (MCS) failure events, which contribute in the occurrence of the top event. These MCS failure events are generally modeled in terms of the exponential or weibull random variables and the Probabilistic Inclusion-Exclusion (PIE) principle \cite{Trivedi_02} is then used to evaluate the exact probability of failure of the given system. However, this method is prone to human errors when it comes to the MCS and failure probability assessment of large safety-critical systems. For instance, in nuclear plants, where a fault tree model involves 50 to 130 levels of logic gates between the top event and the lowest basic events that are contributing to the top event \cite{epstein2005can}. So, there is a possibility, that many of these basic failure events may be overlooked while calculating MCS and thus not further incorporated in the FTA, which may lead to erroneous designs.

The FTA-based computer simulators, such as Relia-Soft \cite{Reliasoft_14} and ASENT Reliability analysis tools \cite{ASENT_14}, provide graphical editors for the construction of FTs and the analysis is carried out by generating samples from the exponential and Weibull random variables that are associated with the events of the FT. These samples are then processed to evaluate the reliability and the failure probability of the complete system using computer arithmetic and numerical techniques. Although, these tools provide a more scalable alternative to the paper-and-pencil proof methods but the computational requirement increases drastically as the size of the FT increases. For example, if there are $q$ terms involves in the MCS of a given FT then the total number of terms in the corresponding PIE principle will be $2q - 1$. In addition, these tools cannot ascertain absolute correctness or error-free analysis because of the involvement of pseudo random numbers and numerical methods and the inherent sampling-based nature of simulation.

Similarly, computer algebra systems (CAS), such as Mathematica \cite{long2000quantification}, provide extensive features for FT-based failure analysis. For instance, the MCS expressions for any given system can be validated with failure distributions, such as Exponential or Weibull, by using symbolic and numerical algorithms. However, due to the presence of these unverified simplification algorithms, the analysis provided by CAS cannot be termed as sound and accurate.

Formal methods can overcome the above-mentioned inaccuracy limitations of the traditional techniques and thus have been used for FTA. The Interval Temporal Logic (ITS), i.e., a temporal logic that supports first-order logic,  has been used, along with the Karlsruhe Interactive Verifier (KIV), for formal FTA of a rail-road crossing \cite{ortmeier2007formal}.
The work presented in \cite{xiang2004fault} describes a deductive method for FT construction, in contrast to the intuitive approach followed in \cite{ortmeier2007formal}, by using the Observational Transition Systems (OTS) \cite{xiang2004fault} and then the formal analysis of this FT is carried out using CafeOBJ \cite{futatsugi2000cafe}, which is a formal specification language with interactive verification support. One of the main limitations of all the above-mentioned formal methods based works is the inability to conduct a probability theoretic FTA. The COMPASS tool-set \cite{bozzano2009compass}, which is developed at RWTH Achen University in collaboration with the European Space Agency (ESA), caters for this problem and supports the formal FTA specifically for aerospace systems using the NuSMV  and MRMC model checkers. However, the scope of these tools is somewhat limited in terms of handling failure analysis of large FTs, due to the inherent state-space explosion problem of model checking, and the fact that the computation of probabilities in these methods involve numerical methods, which compromises the accuracy of the results.

An accurate MCS calculation and exact failure probability assessment in the FTA is very important specially while dealing with safety-critical systems used in domains like transportation, aerospace or medicine. In order to achieve an accurate and precise FTA, we propose to conduct the formal FTA within the sound core of a higher-order-logic theorem prover \cite{harrison_09}. Higher-order logic provides a precise deductive mechanism that can be used to model any mathematically expressive behavior including recursive definitions, random variables, fault tree events, which are the foremost building blocks for modeling FTs. Once the FTs are modeled in higher-order logic, we can deduce an accurate MCS by using formal reasoning based on the set-theoretic foundations. Moreover, FT properties, such as the probability of failure, can be formally verified using interactive theorem provers based on the PIE principles.

The foremost requirement for reasoning about reliability and failure related properties of a system in a theorem prover is the availability of the higher-order-logic formalization of probability theory. Hurd's formalization of measure and probability theories \cite{hurd_02} is a pioneering work in this regard. Building upon this formalization most of the commonly-used continuous random variables \cite{hasan_cade_07} and some reliability theory fundamentals \cite{abbasi_13} have been formalized using the HOL theorem prover. However, Hurd's formalization of probability theory \cite{hurd_02} only supports the whole universe as the probability space. This feature limits its scope in many aspects \cite{mhamdi_11} and one of the main limitations, related to FTA-based analysis, is the nonability to reason about multiple continuous random variables \cite{hasan_cade_07}. Some recent probability theory formalizations \cite{mhamdi_11,holzl_11} allow using any arbitrary probability space that is a subset of the universe and thus are more flexible than Hurd's formalization of probability theory. Particularly, Mhamdi's probability theory formalization \cite{mhamdi_11}, which is based on extended-real numbers (real numbers including $\pm\infty$), has been recently used to reason about the Reliability Block Diagram (RBD)-based analysis of a series pipelines structure \cite{WAhmad_CICM14}, which involves multiple exponential random variables. The current paper is mainly inspired from this development as we use Mhamdi's formalized probability theory \cite{mhamdi_11} for the formalization of all the commonly used FTA gates and the formal verification of their probabilistic properties. Moreover, we have also formally verified the PIE principle, which provides the foremost foundation for formal reasoning about the accurate failure analysis of any FT.

In order to illustrate the effectiveness of the proposed FTA approach, the paper presents a formal failure analysis, by taking a FT model, of a solar array that has been used in the DFH-3 Satellite, which was launched by the People's Republic of China on May 12, 1997 \cite{wu2011reliability}. Solar arrays are one of the most vital components of the satellites because the mission success heavily depends upon the continuous reliable source of power \cite{brandhorst2008space}. Over the last ten years, 12 out of the 117 satellite's solar array anomalies, documented by the Airclaim’s Ascend SpaceTrak database, led to the total satellite failure \cite{SpaceTrack,brandhorst2008space}. Thus the absolute accuracy of the failure analysis of a solar array is a dire need in satellite missions and, to the best of our knowledge,  it is the novelty of the proposed technique to meet this requirement. The satellite's solar array is a mechanical system, which mainly consists of various mechanisms, including: deployable, synchronization, locking and orientation. The FT of the solar array contains the failure events of these mechanisms and their interrelationships regarding the overall system failure. The paper presents the higher-order-logic modeling of this FT and the formal verification of the probability of failure of satellite's solar array system based on the probability of occurrence of the above-mentioned mechanism faults.

\section{Probability Theory in HOL}
In this section, we provide a brief overview of the HOL4 formalization of the probability theory \cite{mhamdi_11}, which we build upon in this paper. Based on the measure theoretic foundations, a probability space is defined as a triple ($\Omega,\Sigma, Pr$), where
 $\Omega$ is a set, called the sample space, $\Sigma$ represents a $\sigma$-algebra of subsets of
$\Omega$, where the subsets are usually referred to as measurable sets, and $Pr$ is a measure with domain
$\Sigma$ and is 1 for the whole sample space. In the HOL4 probability theory formalization \cite{mhamdi_11}, given a probability space $p$, the functions \texttt{space}
and \texttt{subsets} return the corresponding
$\Omega$ and $\Sigma$, respectively. Based on this definition, all the basic probability axioms have been verified. Now, a random variable is a measurable function between a probability space and a measurable space, which essentially is a pair
($S,\mathcal{A}$), where $S$ denotes a set and $\mathcal{A}$ represents a nonempty collection of sub-sets of $S$. A random variable is termed as discrete if $S$ is a set with finite elements and continuous otherwise.

 The cumulative distribution function (CDF) is defined as the probability of the event where a random variable $X$ has a value less than or equal to some value $x$, i.e., $Pr(X \le x)$. This definition characterizes the distribution of both discrete and continuous random variables and has been formalized \cite{WAhmad_CICM14} as follows:

\begin{flushleft}
\label{CDF_def}
\vspace{1pt} \texttt{$\vdash$ $\forall$  p X x. CDF p X x = distribution p X \{y | y $\leq$ Normal x\}
}
\end{flushleft}

\noindent The function \texttt{Normal} takes  a $real$ number as its input and converts it to its corresponding value in the $extended$-$real$ data-type, i.e, it is the $real$ data-type with the inclusion of  positive and negative infinity. The function \texttt{distribution} takes three parameters:  a probability space $p$, a random variable $X$ and a set of $extended$-$real$ numbers and returns  the probability of the given random variable $X$ acquiring  all the values of the given set in probability space $p$.

The unreliability or the probability of failure $F(t)$ is defined as the probability that a system or component will fail by the time $t$. It can be described in terms of CDF, known as the failure distribution function, if the random variable $X$ represent a time-to-failure of the component. This time-to-failure random variable $X$ usually exhibits the exponential or weibull distribution.
%
%
%
%


The notion of mutual independence of $n$ random variables is a major requirement for reasoning about the failure analysis of most of the FT gates. According to this notion, if we have $N$ mutually independent failure events then
\begin{equation}\label{eq1:mutual_indep}
 Pr(\bigcap_{i=1}^{N}L_i) = \prod_{i=1}^{N} Pr(L_i)
\end{equation}
This concept has been formalized as follows \cite{WAhmad_CICM14}:
\begin{flushleft}
 \label{mutual_indep_def}
\vspace{1pt} \small{\texttt{$\vdash$ $\forall$ p L. mutual\_indep p L  =
  $\forall$  L1 \vspace{1pt} n. PERM L L1 $\wedge$\\
\ \ \ \ \ 1 $\leq$ n $\wedge$ n $\leq$ LENGTH L $\Rightarrow$ \\
\ \ \ \ \ \  prob p (inter\_list p (TAKE n L1)) = \\ \ \ \ \ \ \  list\_prod (list\_prob p (TAKE n L1))
}}
\end{flushleft}
\noindent  The function \texttt{mutual\_indep}  accepts a list of events $L$ and probability space $p$ and returns $True$ if the events in the given list are mutually independent in the probability space $p$.  The predicate \texttt{PERM} ensures that its two list arguments  form a permutation of one another. The function \texttt{LENGTH} returns the length of the given list. The function \texttt{TAKE} returns the first $n$ elements of its argument list as a list. The  function \texttt{inter\_list} performs the intersection of all the sets in its argument list of sets  and returns the probability space if the given list of sets is empty. The function \texttt{list\_prob} takes a list of events and returns a list of probabilities associated with the events in the given list of events in the given probability space. Finally, the function \texttt{list\_prod} recursively multiplies all the elements in the given list of real numbers. Using these functions, the function \texttt{mutual\_indep} models the mutual independence condition such that for any 1 or more events $n$ taken from any permutation of the given list $L$, the property $Pr(\bigcap_{i=1}^NL_i) = \prod_{i=1}^NPr(L_i)$ holds.

\section{Formalization of Fault Tree Gates}
In this section, we describe a generic formalization of commonly used FT gates given in Table 1.  Our formalizations are generic in terms of the number of inputs $n$, i.e., our definitions can be used to model arbitrary-input FT gates.
%
\subsection{Formal Definitions of Fault Tree Gates}

If the occurrence of the output failure event is caused by the occurrence of all the input failure events then this kind of behavior can be modeled by using the AND FT gate.
\begin{table}
\centering
\caption{HOL4 Formalization of Fault Tree Gates}
\begin{tabular}{|l |l|}
\hline
Fault Tree Gates & HOL Formalization  \\
\hline
\hline
\parbox[s]{0.5em}{
\includegraphics[width=0.5in]{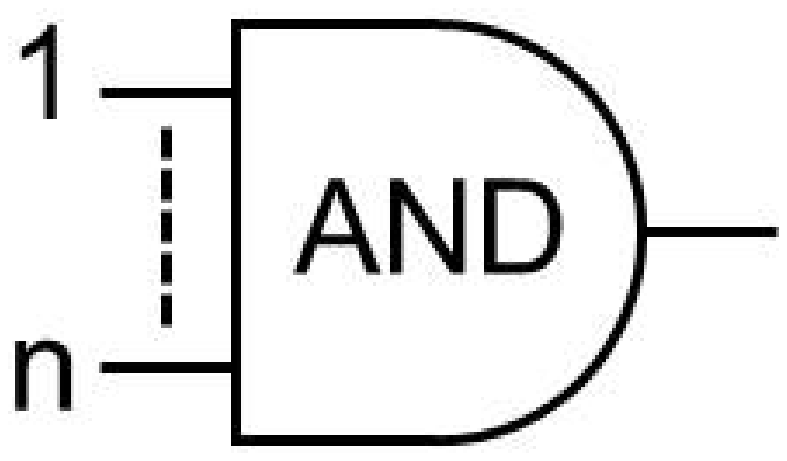}}
& \small{\texttt{$\vdash$ $\forall$ p L. AND\_FT\_gate p L =  inter\_list p L}
}\\

\parbox[c]{0.5em}{
\includegraphics[width=0.5in]{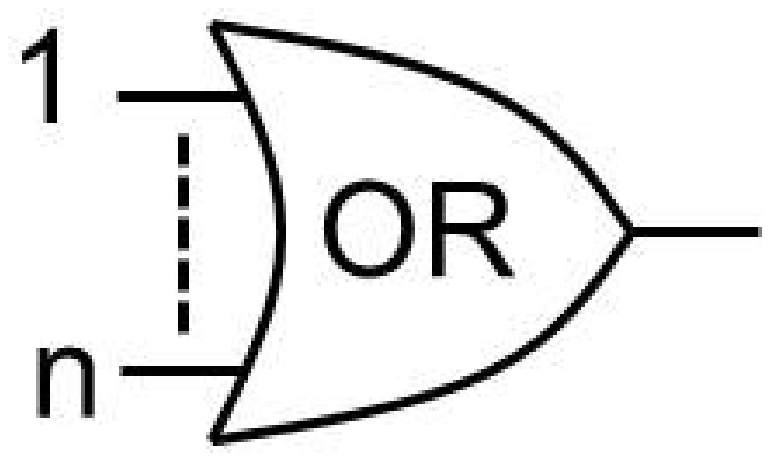}} &  \small{\texttt{$\vdash$ $\forall$ L. OR\_FT\_gate L =  union\_list L
}} \\

\parbox[t]{0.5em}{
\includegraphics[width=0.5in]{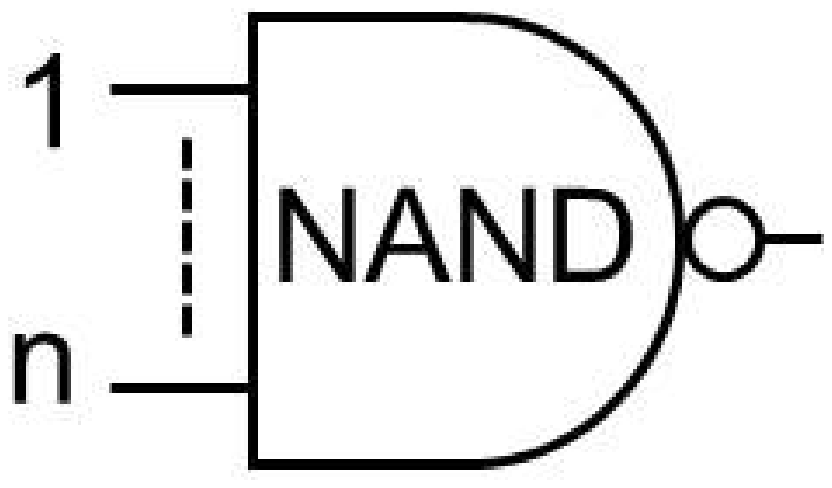}} & \shortstack{\small{\texttt{$\vdash$ $\forall$ p L1 L2. NAND\_FT\_gate p L1 L2 = }} \\ \ \ \ \quad  \small{ \texttt{inter\_list p  (compl\_list p L1)  $\cap$ inter\_list p L2}
}} \\

\parbox[l]{0.5em}{
\includegraphics[width=0.5in]{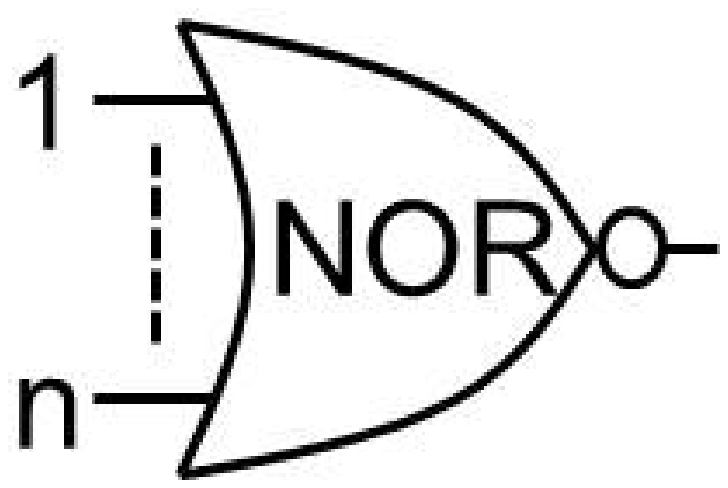}}& \small{\texttt{$\vdash$ $\forall$ p L. NOR\_FT\_gate p L =  p\_space p DIFF (OR\_gate L)}
} \\

\parbox[t]{0.5em}{
\includegraphics[width=0.5in]{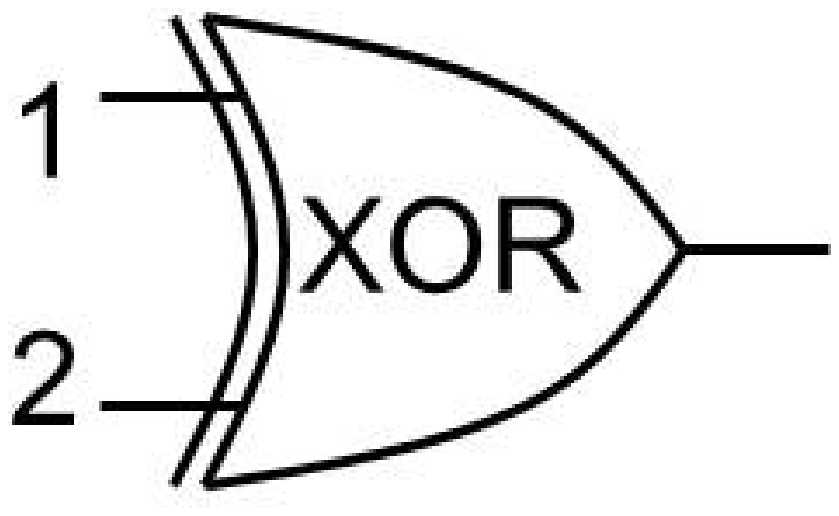}} & \shortstack{\small{\texttt{$\vdash$ $\forall$ p A B. XOR\_FT\_gate p A B =}} \\ \small{\texttt{((p\_space p DIFF A $\cap$ B) $\cup$ (A $\cap$ p\_space p DIFF B))
}}} \\

\parbox[l]{0.5em}{
\includegraphics[width=0.5in]{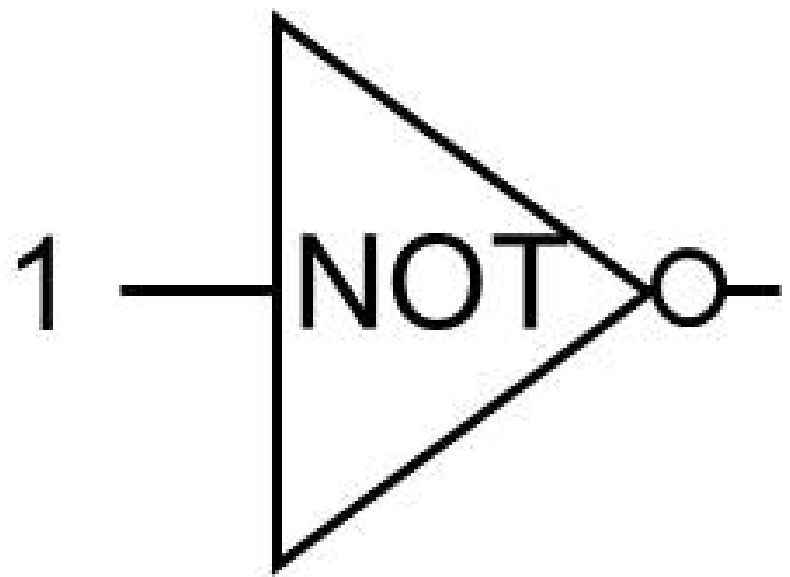}}& \small{\texttt{$\vdash$ $\forall$ p A. NOT\_FT\_gate p A =  (p\_space p DIFF A)
}} \\
\hline
\end{tabular}
\end{table}
The function \texttt{AND\_FT\_gate}, given in Table 1, models this behavior as it accepts an arbitrary probability space $p$  and returns the intersection of input failure events, given in the list $L$,  by using the recursive function \texttt{inter\_list}.
%

In the OR FT gate, the occurrence of the output failure event depends upon the occurrence of any one of its input failure event.
The  function \texttt{OR\_FT\_gate}, given in Table 1, models this behavior as it returns the union of the input failure list $L$ by using the recursive function \texttt{union\_list}. The NOR FT gate can be viewed as the complement of the OR FT gate and its output failure event occurs if none of the input failure event occurs.
The NAND FT gate models the behavior of the occurrence of an output failure event when at least one of the failure events at its input does not occur. This type of gate is used in FTs when the non-occurrence of the failure event in conjunction with the other failure events causes the top failure event to occur. This behavior can be expressed as the intersection of  complementary and normal events {\cite{international2006iec}, where the complementary events model the non-occurring failure events and the normal events model occurring failure events. It is important to note that the behavior of the NAND FT gate is usually not captured by the complement of the AND FT gate in the FTA literature \cite{international2006iec}.
%
The function \texttt{NAND\_FT\_gate} accepts a probability space $p$ and two list of failure events $L1$ and $L2$. The function returns the intersection of non-occurring failure events, which in turn is modeled by passing the list of failure events $L1$ to the recursive function \texttt{compl\_list}, and occurring failure events, which are given in the list $L2$, by utilizing the recursive function \texttt{inter\_list}. The function \texttt{compl\_list} returns a list of events such that each element of this list is the difference between the probability space $p$ and the corresponding element of the given list.
%
%
%

The output failure event occurs in the 2-input XOR FT gate if only one, and not both, of its input failure events occur. The HOL representation of the behaviour of the XOR FT gate is presented in Table 1. The function \texttt{NOT\_FT\_gate} accepts an arbitrary failure event $A$ along with probability space $p$ and returns the complement to the probability space $p$ of the given input failure event $A$.
%
%
%
%
%
%
%
\subsection{Formal Verification of Failure Probability of Fault Tree Gates}

The function \texttt{AND\_FT\_gate}, given in Table 1, can be used to evaluate the failure probability of the output failure event of the AND FT gate. If $A_{i}$ represents the $i^{th}$ failure event with failure probability $F_{i}$ at time $t$ among the $n$ mutually independent failure events of the AND FT gate then the generic mathematical expression for the failure probability of a $n$-input AND FT gate is as follows:
\begin{equation}\label{eq3:and_gate}
      F_{AND\_gate}(t) = Pr (\bigcap_{i=2}^{N}A_{i}(t))
      = \prod_{i=2}^{N}F_{i}(t)
 \end{equation}

\noindent We formally verified this expression as the following theorem in HOL4:
\begin{flushleft}
\small {\texttt{\bf{Theorem 1: }}} \label{AND_gate_THM}
\vspace{1pt} \small{\texttt{$\vdash$ $\forall$ p L. prob\_space p $\wedge$\\
\ \ \ \ \ 2 $\leq$ LENGTH L $\wedge$ mutual\_indep p L $\Rightarrow$ \\
\ \ \ \ \  \ \ \ \  (prob p (AND\_gate p L) = list\_prod (list\_prob p L))
}}
\end{flushleft}
The first assumption ensure that $p$ is a valid probability space based on the probability theory in HOL4  \cite{mhamdi_11}. The next two assumptions guarantee that the list of failure events must have at least two failure event and the failure events are mutually independent, respectively. The conclusion of the theorem represents Equation (\ref{eq3:and_gate}). The proof of Theorem 1 is primarily based on some probability theory axioms and the mutual independence definition.

Similarly, if $A_{i}$ represents the $i^{th}$ with failure event failure probability $F_{i}$ at time $t$ among the $n$ mutually independent failure events of an OR FT gate then its failure probability expression is as follows:
\begin{equation}\label{eq4:OR_gate}
      F_{OR\_gate}(t) = Pr (\bigcup_{i=2}^{N}A_{i}(t))
      = 1 - \prod_{i=2}^{N}(1 - F_{i}(t))
 \end{equation}

In order to formally verify the above equation, we first formally verify the following lemma that provides an alternate expression for the failure probability of an OR FT gate in terms of the failure probability of an AND FT gate:
\begin{flushleft}
\small{\texttt{\bf{Lemma 1: }}}
\vspace{1pt} \small{\texttt{$\vdash$ $\forall$ L p. (prob\_space p) $\wedge$ \\
($\forall$ x'. MEM x' L  $\Rightarrow$  x' $\in$ events p) $\Rightarrow$ \\
\ \     (prob p (OR\_gate L) = \\
\ \  1 - prob p (AND\_gate p (compl\_list p L))
}}
\end{flushleft}

\noindent Now, we can formally verify Equation (\ref{eq4:OR_gate}) in HOL4 as follows:
\begin{flushleft}
\small{\texttt{\bf{Theorem 2: }}} \label{OR_gate_THM}
\vspace{1pt}\small{ \texttt{$\vdash$ $\forall$ p L. (prob\_space p) $\wedge$\\
(2 $\leq$ LENGTH L) $\wedge$ (mutual\_indep p L) $\wedge$ \\
\ ($\forall$ x'. MEM x' L $\Rightarrow$  x' $\in$ events p) $\Rightarrow$ \\
\ \   (prob p (OR\_gate L)  = \\
\ \    1 - list\_prod (one\_minus\_list (list\_prob p L)))
}}
\end{flushleft}
\noindent Where the function \texttt{one\_minus\_list} accepts a list of $real$ numbers $[x_{1}, x_{2}, \cdots, x_n]$ and returns the list of $real$ numbers such that each element of this list is 1 minus the corresponding element of the given list, i.e., $[1-x_1, 1-x_2, \cdots, 1-x_n]$.
The proof of Theorem 2 is primarily based on Lemma 1 and Theorem 1 along with the fact that given the list of $n$ mutually independent events, the complement of these $n$ events are also
mutually independent.

Similarly, we also verified the failure probability theorems for other FT gates, given in Table 1, and the corresponding mathematical expressions and theorems are given in Table 2. All these results are verified under the same assumptions as the ones used in Theorems 1 and 2.

\begin{table}
\centering
\caption{Probability of Failure of Fault Tree Gates}
\begin{tabular}{|l|l|}
\hline
Fault Tree Gates & Theorem's Conclusion \\
\hline
\hline
$\!\begin{aligned}[t]
 F_{NOR}(t)
    & = 1 - F_{OR}(t)  \\
    &= \prod_{i=2}^{N}(1 - F_{i}(t))
    \end{aligned}$  &
 $\!\begin{aligned}[t] & \small{\texttt{ (prob p (NOR\_FT\_gate p L)  =}} \\ &
\ \  \small{\texttt{list\_prod (one\_minus\_list} }\\ & \qquad \small{\texttt{(list\_prob p L)))
}}  \end{aligned}$ \\
\hline
$\!\begin{aligned}[t]
F_{NAND}(t) & =  Pr (\bigcap_{i=2}^{k}\overline A_{i}(t) \cap \bigcap_{j=k}^{N}A_{i}(t)) \\ &= \prod_{i=2}^{k}(1 - F_{i}(t)) *\prod_{j=k}^{N}(F_{j}(t))\end{aligned}$  &  $\!\begin{aligned}[t]& \small{\texttt{(prob p (NAND\_FT\_gate p L1 L2)  = }}\\ &
\ \small{\texttt{list\_prod ((list\_prob p}} \\ & \ \ \ \ \small{\texttt{(compl\_list p L1))) *}}\\ & \small{\texttt{ list\_prod (list\_prob p L2))}} \end{aligned}$ \\
\hline
$\!\begin{aligned}[t]
 F_{XOR}(t)&= Pr(\bar{A}(t)B(t) \cup A(t)\bar{B}(t)) \\ &= (1- F_{A}(t))F_{B}(t) + \\ & \qquad F_{A}(t)(1- F_{B}(t))\end{aligned}$  & $\!\begin{aligned}[t]& \texttt{(prob p (XOR\_FT\_gate p A B)  =}\\ & \texttt{(1- prob p A)*prob p B + }\\ &\quad \texttt{prob p A*(1 - prob p B)} \end{aligned}$ \\
\hline
$\!\begin{aligned}[t]
 F_{NOT}(t)&= Pr(A(t)) \\ &=(1 - F_{A}(t))\end{aligned}$  & $\!\begin{aligned}[t] & \texttt{prob p (NOT\_FT\_gate p A)} = \\ & \texttt{(1 - prob p A)}\end{aligned}$  \\
\hline
\end{tabular}
\end{table}

The proof script \cite{waqar_rbd_nfm_15}  of the above-mentioned formalization is composed of 4000 lines of HOL script and took about 200 man-hours. The main outcome of this exercise is that the definitions, given in Table 1, can be used to capture the behavior of most of the FTs in higher-order logic and the Theorems of Table 2 can then be used in conjunction with the formalization of the PIE principle, explained  next, to formally verify the corresponding failure probabilities.
%
%
%
 \section{Formalization of Probabilistic Inclusion-Exclusion Principle }
 The probabilistic inclusion-exclusion principle (PIE) forms an integral part of the reasoning involved in verifying the failure probability of a FT. In FTA, firstly all the basic fault events are identified that can cause the occurrence of the system failure event. These fault events are then combined to model the overall fault behavior of the given system by using the fault gates. These combinations of basic failure events, called cut sets, are then reduced
to  minimal cut sets (MCS) by using some set-theory rules, such as idempotent, associative and commutative \cite{halmos1960naive}. At this point, the PIE principle is used to evaluate the overall failure probability of the given system  based on the MCS events.

If $A_{i}$ represent the $i^{th}$ basic failure event or a combination of failure event then the failure probability of the given system can be expressed in terms of the probabilistic inclusion-exclusion principle as follows:

 \begin{equation}\label{PIE}
\mathbb{P} (\bigcup_{i=1}^n A_i)  = \sum_{t \neq \{\}, t\subseteq\{1,2,\ldots,n\}}(-1)^{|t|+1} \mathbb{P} (\bigcap_{j\in t} A_j)
 \end{equation}

\noindent The above equation can be formalized in HOL4 is as follows:
 \begin{flushleft}
\small{\texttt{\bf{Theorem 3: }}} \label{PIE_THM}
\vspace{1pt} \small{\texttt{$\vdash$ $\forall$ p L1 L2. prob\_space p $\wedge$ \\ ($\forall$ x. MEM x L $\Rightarrow$ x $\in$ events p) $\Rightarrow$ \\
\qquad     (prob p (union\_list L) = \\ \qquad
      sum\_set \{t | t  $ \subseteq $ set L $ \wedge $ t $ \neq $ \{\} \}\\ \qquad \qquad
        ($ \lambda $t. -1 pow (CARD t + 1) * prob p (BIGINTER t)))
}}
\end{flushleft}

\noindent The assumptions of the above theorem are the same as the ones used in Theorem 1. The function \texttt{sum\_set} takes an arbitrary set $s$ with element of type $\alpha$ and a real-valued function $f$. It recursively sums the return value of the function $f$, which is applied on each element of the given set $s$. In the above theorem, the set $s$ is represented by the term $\{x|C(x)\}$ that contains all the values of $x$, which satisfy condition $C$. Whereas, the $\lambda$ abstraction function \texttt{($ \lambda $t. -1 pow (CARD t + 1) * prob p (BIGINTER t))} models $(-1)^{|t|+1} \mathbb{P} (\bigcap_{j\in t} A_j)$, such that the functions \texttt{CARD} and \texttt{BIGINTER} return the number of elements and the intersection of all the elements of the given set, respectively. Thus, the conclusion of the theorem represents Equation (\ref{PIE}).

The formal reasoning about Theorem 3 is based upon the following lemma:

\begin{flushleft}
\small{\texttt{\bf{Lemma 2: }}} \label{PIE_lemma_THM}
\vspace{1pt} \texttt{$\vdash$ $\forall$ P. ($\forall$ n. ($\forall$ m. m < n $\Rightarrow$ P m) $\Rightarrow$ P n) $\Rightarrow$ $\forall$ n. P n
}
\end{flushleft}

\noindent Where $n$ in our case is the length of the list $L$ and $m$ represent another list whose length is less then the length of the list $L$. The predicate $P$ represents the conclusion of  Theorem 3. The above property brings an important hypothesis in the assumption list, which has the same form as that of the conclusion of Theorem 3. Then, by utilizing induction and some properties of the function \texttt{sum\_set} along with some fundamental axioms of probability, we can verify Theorem 3.

The proof script \cite{waqar_rbd_nfm_15}  for Theorem 3 is composed of 1000 lines of HOL code and involved 50 man-hours of proof effort. To the best of our knowledge, this is the first formal verification of the probabilistic inclusion exclusion principle, which, besides being used in FTA, is a widely used mathematical result in analyzing various bio-informatics \cite{Todor15062014} and telecommunication \cite{apploni} systems.
\section{Application: Satellite's Solar Array}
The solar arrays used in satellite missions are usually in a folded position during the launch phase \cite{wu2011reliability}. Once the satellite is deployed in the corresponding orbit then the solar arrays are unfolded and the goal is to  keep them oriented towards the sun all the time to maximize the power generation for the satellite \cite{wu2011reliability}. The faults in the solar array are mainly caused by the mechanical components that drive these mechanisms associated with the driving, deployment, synchronization, locking and orientation. For example, the solar array is usually driven by using a torsion spring \cite{wu2011reliability}. Whereas, the closed cable loop (CCL) and the stepping or servo motors are used during the synchronization and orientation phases \cite{wu2011reliability}. A FT can thus be constructed by considering the faults in these mechanical components, which are the fundamental causes of satellite' solar array mechanisms failure The FT for the solar array of the DFH-3 Satellite that was launched by the People's Republic of China on May 12, 1997 \cite{jianing2011reliability} is depicted in Figure 1 and we formally analyze this FT in this paper.

The failure events, \textit{A, B, C, D}  in Figure 1, represent the failures in the unlock mechanism, deployment process, locking process and orientation process, respectively. Whereas, the failure event \textit{E} represents the failures in the corresponding mechanical parts of the system. These failure events are combined either by using the OR or AND FT gates by considering the behavior of the faults.

In order to formalize the solar array FT of Figure 1, we first present the formal modeling of list of failure events that are associated with each corresponding fault of the solar array FT.
\begin{flushleft}
\texttt{\bf{Definition 1: }}
\label{list_RV_def}
\vspace{1pt} \small{\texttt{$\vdash$ $\forall$ p x. fail\_event\_list p [] x = [] $\wedge$ \\
\ \ $\forall$ p x h t. fail\_event\_list p (h::t) x = \\
\ \ \ \ \ PREIMAGE h \{y | y $\leq$ Normal x \} $\cap$ p\_space p :: \\ \quad \ \ \ \ fail\_event\_list p t x
}}
\end{flushleft}
\noindent The function \texttt{fail\_event\_list} accepts a probability space $p$, a list of random variables, representing the failure time of individual components, and a real number $x$, which represents the time index at which the failure of the component occurs. It returns a list of events, representing the failure of all the individual components at time $x$.
The formal definitions of FT gates, given in Section 3, along with Definition 1 can be utilized to formally represent the FT of satellite's solar array in terms of its cut-set failure events. The HOL4 formalization of satellite's solar array FT is as follows:
\begin{flushleft}
\texttt{\bf{Definition 2: }}
 \label{solar_FT_CS_def}
\vspace{1pt} \small{\texttt{$\vdash$ $\forall$ p x1 x2 x3  x4 x5 x6 x7 x8 x9 x10 x11 \\ \qquad \qquad \qquad \qquad \qquad \qquad x12 x13 x14 t. \\
     Solar\_FT p x1 x2 x3 x4 x5 x6 x7 x8 x9 x10 x11 x12 x13 x14 t = \\
    OR\_FT\_gate
       [OR\_FT\_gate (fail\_event\_list p [x1; x2] t); \\ \qquad \qquad \quad \ \
        OR\_FT\_gate
          [OR\_FT\_gate (fail\_event\_list p [x3; x4] t);\\ \qquad \qquad \qquad \qquad AND\_FT\_gate p (fail\_event\_list p [x5; x6] t);
          OR\_FT\_gate (fail\_event\_list p [x3; x7; x8] t)];\\ \qquad \qquad \qquad OR\_FT\_gate (fail\_event\_list p [x3; x9] t); \\
\qquad \qquad \qquad        OR\_FT\_gate (fail\_event\_list p [x10; x11] t);\\
\qquad \qquad \qquad        OR\_FT\_gate [PREIMAGE x12 \{y | y $\leq$ Normal t \}; \\ \qquad \qquad \qquad \qquad \qquad \quad PREIMAGE x13 \{y | y $\leq$ Normal t \}; \\ \qquad \qquad \qquad \qquad \qquad \quad OR\_FT\_gate (fail\_event\_list p[x3; x14]t)]]
}}
\end{flushleft}
\noindent Where the random variables $x1-x14$ model the time-to-failure of the solar array processes and components as depicted in Figure 1.
\begin{figure}[!ht]\label{logic-gates}
\centering
\includegraphics[width= 0.9\textwidth]{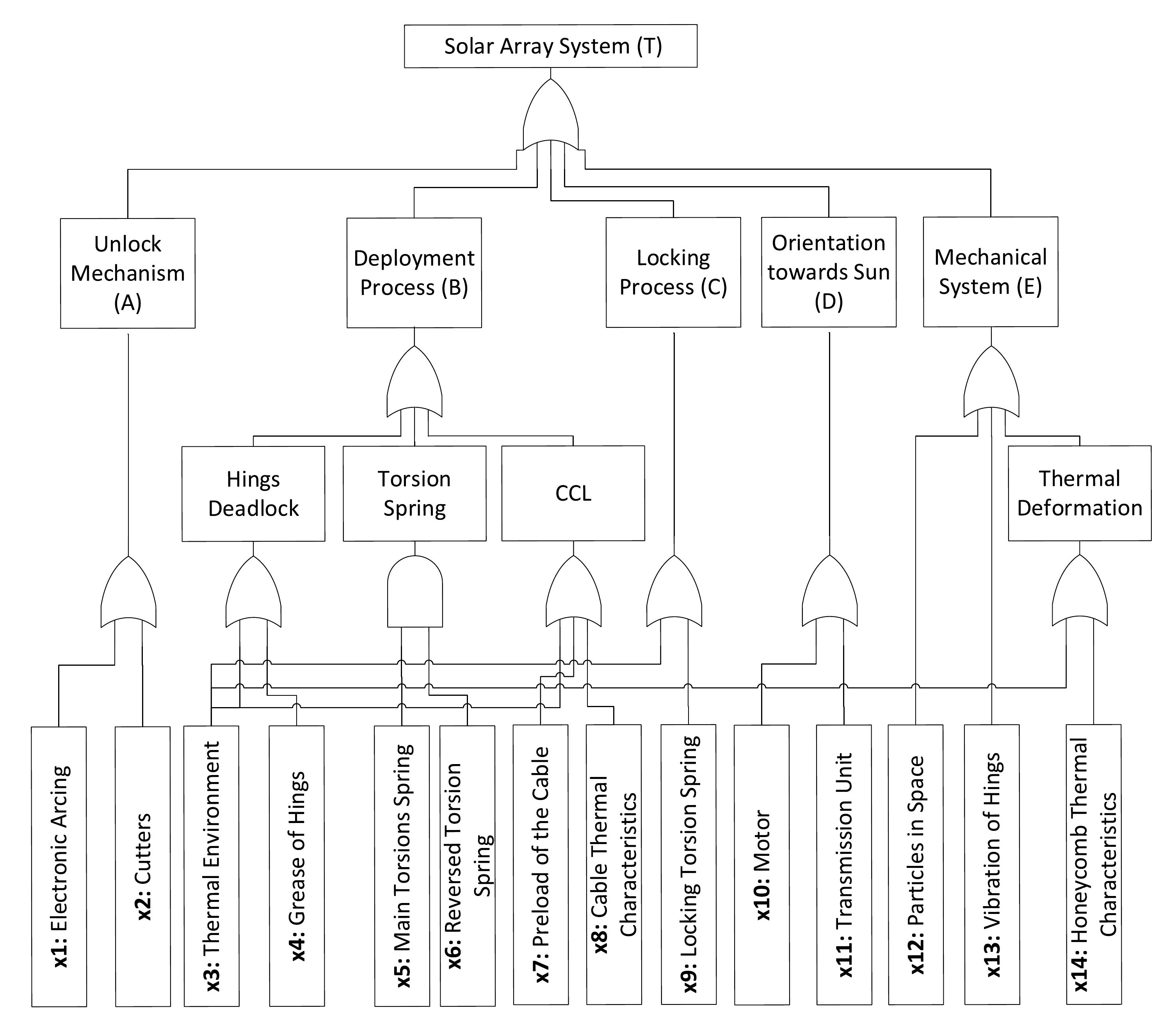}\caption{FT of the Solar Array of the DFH-3 Satellite \cite{wu2011reliability}}
\end{figure}
 However, the cut-set failure events in the above definition is not minimal \cite{wu2011reliability}, i.e., there are some redundant failure events. For example, $x3$ is part of more than one OR FT gates. These kind of redundant failure events can be removed by verifying an accurate equivalent but reduced representation, i.e., the MCS, by using set theory laws, like idempotent, commutative and associative, as follows:
 \begin{flushleft}
\small{\texttt{\bf{Lemma 2: }}} \label{lemma_solar cell_THM}
\vspace{1pt} \small{\texttt{$\vdash$ $\forall$ p x1 x2 x3 x4 x5 x6 x7 x8 x9 x10 x11 x12 x13 x14 t.\\
\qquad \qquad     prob\_space p $\Rightarrow$ \\
     (Solar\_FT p x1 x2 x3 x4 x5 x6 x7 x8 x9 x10 x11 x12 x13 x14 t = \\
      OR\_FT\_gate
        [OR\_FT\_gate  (fail\_event\_list p [x1; x2; x3; x4] t);\\ \qquad \qquad \qquad AND\_FT\_gate  p (fail\_event\_list p [x5; x6] t);\\
 \qquad \qquad \qquad        OR\_FT\_gate \\  \qquad \qquad \qquad  \ (fail\_event\_list p [x7; x8; x9; x10; x11; x12; x13; x14]t)])
}}
\end{flushleft}

We consider that random variables, associated with the failure events of the solar array FT, exhibit the exponential distribution, which can be formalized in HOL4 as follows:
\begin{flushleft}
\small{\texttt{\bf{Definition 3: }} \label{Exponential_distribution_def}
\vspace{1pt} \texttt{$\vdash$ $\forall$ p X l. exp\_dist p X l = \\ \ \ \  $\forall$ x.  (CDF p X x =  if 0 $\leq$ x then 1 - exp (-l * x) else 0)
}}
\end{flushleft}

\noindent The function \texttt{exp\_dist} guarantees that the CDF of the random variable $X$ is that of an
exponential random variable with a failure rate $l$ in a probability space $p$. We classify a list of exponentially distributed random variables based on this definition as follows:
\begin{flushleft}
\small{\texttt{\bf{Definition 4: }} \label{list_of exponential_distribution_function_def}
\vspace{1pt} \small{\texttt{$\vdash$ $\forall$ p L. list\_exp p [] L = T $\wedge$\\
\ \ $\forall$ p h t L.  list\_exp p (h::t) L = \\ \ \qquad \qquad \qquad exp\_dist p (HD L) h $\wedge$ list\_exp p t (TL L)
}}}
\end{flushleft}

\noindent The function \texttt{list\_exp} accepts a list of failure rates, a list of random variables $L$ and
a probability space $p$. It guarantees that all elements of the list $L$ are exponentially distributed with the corresponding failure rates, given in the other list, within the probability space $p$. For this purpose, it utilizes the list functions \texttt{HD} and \texttt{TL}, which return the \emph{head} and \emph{tail} of a list, respectively.
Now, the failure probability of satellite's solar array can be verified as the following theorem:

\begin{flushleft}
\small{\texttt{\bf{Theorem 4: }}} \label{solar cell_THM}
\vspace{1pt} \small{\texttt{$\vdash$ $\forall$ p x1 x2 x3 x4 x5 x6 x7 x8 x9 x10 x11 x12 x13 x14 t c1 c2 c3 c4 c5 c6 c7 c8 c9 c10 c11 c12 c13 c14.
\\ (0 $\le$ t) $\wedge$ (prob\_space p) $\wedge$ \\  ($\forall$ x'. MEM x'
(fail\_event\_list p \\ \ ([x1; x2; x3; x4; x5; \\ \ \quad x6; x6; x7; x8; x9; x10; x11; x12; x13; x14]) t)) $\Rightarrow$ x' $\in$ events p)
  $\wedge$  \\
(mutual\_indep p ((fail\_event\_list p \\ \ \ ([x1; x2; x3; x4; x5; x6; x7; x8; x9; x10; x11; x12; x13; x14])  x)))  $\wedge$ \\ list\_exp p \\ \ ([c1; c2; c3; c4; c5; c6; c7; c8; c9; c10; c11; c12; c13; c14]) \\ \quad \quad ([x1; x2; x3; x4; x5; x6; x7; x8; x9; x10; x11; x12; x13; x14]) $\Rightarrow$ \\
\quad        (prob p
          (Solar\_FT p  \\ \quad \quad x1 x2 x3 x4 x5 x6 x7 x8 x9 x10 x11 x12 x13
             x14 t ) = \\  \quad  (1 -  (exp -(t*(list\_sum [c1;c2;c3;c4])))) + \\ \qquad  list\_prod(one\_minus\_exp t [c5;c6;c7]) + \\ \qquad  (1 -  (exp -(t*(list\_sum \\ \qquad  \quad [c7; c8; c9; c10; c11; c12; c13; c14])))) - \\ \qquad  (1 - list\_prod(one\_minus\_exp\_prod t \\ \qquad \quad [[c1;c5;c6];[c2;c5;c6];[c3;c5;c6];[c4;c5;c6]])) - \\ \qquad (1 -  (exp -(t*(list\_sum [c1;c2;c3;c4])))) * \\ \qquad \quad (1 -  (exp -(t*(list\_sum \\ \qquad  \qquad [c7; c8; c9; c10; c11; c12; c13; c14])))) - \\ \qquad (1 - list\_prod(one\_minus\_exp\_prod t \\ \qquad  \quad [[c5;c6;c7];[c5;c6;c8];[c5;c6;c9];[c5;c6;c10];\\ \qquad \qquad  [c5;c6;c11];[c5;c6;c12];[c5;c6;c13];[c5;c6;c14]])) + \\ \qquad (1 - list\_prod(one\_minus\_exp\_prod t \\ \qquad  \quad [[c1;c5;c6];[c2;c5;c6];[c3;c5;c6];[c4;c5;c6]])) * \\ \qquad  \qquad (1 -  (exp -(t* \\ \qquad  \qquad \quad (list\_sum [c7; c8; c9; c10; c11; c12; c13; c14])))))
}}
\end{flushleft}

\noindent The first assumption ensures the variable \texttt{t} that models time can acquire positive values only. The second assumption ensure that \texttt{p} is a valid probability space based on the probability theory in HOL4 \cite{mhamdi_lebsegue}. The next two assumptions ensure that the events corresponding to the failures modeled by the random variables \texttt{x1} to \texttt{x14} are valid events from the probability space \texttt{p} and they are mutually exclusive. Finally, the last assumption characterizes the random variables \texttt{x1} to \texttt{x14} as exponential random variables with failure rates \texttt{c1} to \texttt{c14}, respectively. The conclusion of the Theorem 4 represents the failure probability of the given solar array in terms of the failure rates of its components as follows:

\small
\begin{equation}\label{failure_solar_Eq}
\begin {split}
  &(1 - e^{-(c1+c2+c3+c4)t}) + \prod_{i=5}^{6} (1-e^{-(c_{i}t)}) + \\ & (1 - e ^{-(c7+c8+c9+c10+c11+c12+c13+c14)t}) - (1 - \prod_{i=1}^{4}(1 - \prod_{j=5}^{6}[(1-e^{-c_{i}t})(1-e^{-c_{j}t})])) -\\
  &(1 - e^{-(c1+c2+c3+c4)t})*(1-e^{-(c7+c8+c9+c10+c11+c12+c13+c14)t}) -\\
  &(1 - \prod_{i=7}^{14}(1 - \prod_{j=5}^{6}[(1-e^{-c_{i}t})(1-e^{-c_{j}t})])) + \\
  &(1 - \prod_{i=1}^{4}(1 - \prod_{j=5}^{6}[(1-e^{-c_{i}t})(1-e^{-c_{j}t})])) * (1 - e ^{-(c7+c8+c9+c10+c11+c12+c13+c14)t})
  \end{split}
\end{equation}
\normalsize

\noindent where the  function \texttt{exp} represents a exponential function, the function \texttt{list\_sum} is used to sum all the element of the given list of failure rates, the function \texttt{one\_minus\_exp} accepts a list of failure rates and returns a one minus list of exponentials and the function \texttt{one\_minus\_exp\_prod}  accepts a two dimensional list of  failure rates and returns a list with one minus product of one minus exponentials of every sub-list. For example, \texttt{one\_minus\_exp\_prod}$[[c1; c2; c3]; [c4; c5]; [c6; c7; c8]]$ $ x = [1 - ((1 - e^ {-(c1)x})*(1 - e^{ -(c2)x})*(1 - e^{ -(c3)x}));$
$(1 - (1 - e^{ -(c4)x}) * (1 - e^ {-(c5)x})); (1 - (1 - e^ {-(c6)x}) * (1 - e^{ -(c7)x}) *(1 - e^ {-(c8)x})) ]$.

 The proof of the above theorem utilizes the failure probabilities of AND and OR FT gates, given in Table 2, along with Lemma 2 and Theorem 3 and some fundamental facts and axioms of probability theory. Due to the universally quantified variables in Theorem 3, the proof of Theorem 4 is quite straight-forward (about 800 lines of HOL code)  as compared to that of Theorem 3. The distinguishing features of the formally verified Theorem 4 includes its generic nature, i.e., all the variables are universally quantified and thus can be specialized to obtain the failure probability for any given failure rates, and its guaranteed correctness due to the involvement of a sound theorem prover in its verification, which ensures that all the required assumptions for the validity of the result are accompanying the theorem.

A fuzzy reasoning Petri Net (FRPN), which is a combination of fuzzy logic \cite{zadeh1997toward} and Petri Nets \cite{peterson1981petri}, based failure analysis for the above-mentioned solar array is presented in \cite{wu2011reliability}. In this work, the FT of Figure 1 is first represented as a Petri Net such that the gates are represented by transitions and the failure events are modeled as places. The possibility of fault occurrence is then evaluated by using fuzzy degree of truth on the basis of petri nets transitions. However, the truth degree values evaluated using these FRPN models cannot be regarded as precise and sound as the formally verified expression using the HOL theorem prover due to the involvement of numerical techniques and pseudo randomness. On the other hand, our analysis result, i.e., Theorem 4, is based on a probability theoretic formal reasoning, verified in a sound theorem prover and is valid for all possible values of the failure rates. These features constitute the main motivations of the work presented in this paper.

\section{Conclusion}

The accuracy of failure analysis is a dire need for safety and mission-critical applications, where an incorrect failure analysis may lead to disastrous situations including the loss of human lives or heavy financial setbacks. In this paper, we presented an accurate FTA approach, based on higher-order-logic theorem proving, to tackle the analysis of such critical systems. In particular the paper presents a formalization of commonly used FT gates and the PIE principle, which are the foremost foundations for formal reasoning about FTA within a sound core of theorem prover. As a case-study, the paper also presents the formal failure analysis of a satellite's solar array.

Building upon the results, presented in this paper, other FT gates, such as priority AND and voting OR gate, can also be formally modeled and thus the scope of FTA-based formal reliability analysis \cite{volkanovski2009application} can be further enhanced. Some interesting real-world applications that can benefit from our work include transportation systems \cite{huang2000fuzzy}, healthcare systems \cite{hyman2008fault} and avionics \cite{lefebvre2007diagnostic}. Moreover, we also plan to further facilitate the formal FT-based failure analysis by incorporating the automatic simplification capabilities of CAS, such as Mathmatica, for MCS calculation. This obtained MCS can then be validated within the sound environment of the HOL theorem prover.

\small
\bibliographystyle{splncs}
\bibliography{biblio}
\end{document}